\documentclass[letterpaper,english,preprint]{revtex4-1}
\usepackage[T1]{fontenc}
\usepackage[latin1]{inputenc}
\usepackage{amsmath}
\usepackage{amssymb}
\usepackage{graphicx}

\makeatletter

\providecommand{\tabularnewline}{\\}

\usepackage{graphicx}

\usepackage{graphicx}

\makeatother

\usepackage{babel}
\begin{document}

\title{A Unified Strouhal-Reynolds Number Relationship for Laminar Vortex
Streets Generated by Different Shaped Obstacles}

\author{Ildoo Kim}

\email{ildoo_kim@brown.edu}

\affiliation{School of Engineering, Brown University, Providence, RI 02906}

\affiliation{Department of Physics and Astronomy, University of Pittsburgh, Pittsburgh,
PA 15260}

\author{X.L. Wu}

\affiliation{Department of Physics and Astronomy, University of Pittsburgh, Pittsburgh,
PA 15260}

\date{August 12, 2015}
\begin{abstract}
A new Strouhal-Reynolds number relationship, $St=1/(A+B/Re)$, has
been recently proposed based on observations of laminar vortex shedding
from circular cylinders in a flowing soap film. Since the new $St$-$Re$
relation was derived from a general physical consideration, it raises
the possibility that it may be applicable to vortex shedding from
bodies other than circular ones. The work presented herein provides
experimental evidence that this is the case. Our measurements also
show that in the asymptotic limit ($Re\rightarrow\infty$), $St_{\infty}=1/A\simeq0.21$
is constant independent of rod shapes, leaving $B$ the only parameter
that is shape dependent. 
\end{abstract}


\maketitle
When a flowing fluid encounters an obstacle, two staggered rows of
vortices form downstream. This so-called von K\'{a}rm\'{a}n vortex
street has been studied by scientists for many years \cite{Strouhal1878,Rayleigh1915,Birkhoff1953,Roshko1954,Williamson1996rev},
but our basic understanding of vortex wake formation, its stability,
and evolution remains incomplete. At the heart of the problem is why
and how vorticity, which is created in the boundary layer and discharged
into a bulk of fluid, self-organizes into spatiotemporally periodic
patterns. In this paper, we wish to address this issue by studying
vortex shedding and street formation using rods of different geometric
cross-sectional areas but with their aspect ratios close to unity.
The experiment is conducted in freely flowing soap films that strongly
suppresses instabilities and turbulence, which are commonly encountered
in three dimensional (3D) fluids when Reynolds number $Re$ is large.
Our quasi 2D experiments therefore allow \emph{laminar vortex streets}
to be studied over a broad range of $Re$. 

In an earlier paper, we showed that a $St-Re$ relation can be derived
based on simple observations of vortex streets beneath a circular
rod \cite{Roushan2005a}. Specifically, if the flow speed $U$ is
held constant and $Re$ is varied by changing the diameter $D$ of
the rod, the experiment shows that the wavelength of the vortex street
$\lambda$ is a linear function of $D$, $\lambda=\lambda_{0}+\alpha D$,
that spans the entire range of $Re$ ($10-3\times10^{3}$) in the
measurement, where $\lambda_{0}\simeq0.1\,{\rm cm}$ and $\alpha\simeq4$
are constant. Another simplifying feature observed in the experiment
is that over the same span of $Re$, vortex street travels at a constant
speed $v_{st}$ relative to the background flow $U$ so that $c(\equiv v_{st}/U)\simeq0.8$
remains constant for different $D$. Since the laminar vortex street
represents a single global mode of fluid dynamics, its frequency must
satisfy $f=v_{st}/\lambda$. Using the definition of $St\left(\equiv fD/U\right)$,
it follows immediately that $St=1/\left(A+B/Re\right)$, or more concisely
\begin{equation}
\frac{1}{St}=A+B\cdot\frac{1}{Re},\label{eq:Roushan-Wu}
\end{equation}
where $A=\alpha/c$ and $B=\lambda_{0}U/(c\nu)$ are constant. 

Eq. \eqref{eq:Roushan-Wu} works well for measurements using circular
rods in 3D fluids as well as in 2D soap films as discussed in Ref.
\cite{Roushan2005a}. Unlike previously proposed $St-Re$ relation
that is either applicable to vortex streets near an onset \cite{Provansal1987,Monkewitz1988,Chomaz1988}
or very far from it \cite{Birkhoff1953}, the remarkable fact is that
Eq. \eqref{eq:Roushan-Wu} is applicable to both low and high $Re$.
The robustness of this relationship is a testament of the fact that
our empirical approach is capable of capturing important features
of vortex street behind a bluff body. The purpose of the current research
is therefore two folds: (i) to further explore this approach by investigating
vortex streets created by (blunt) bodies of different shapes, and
(ii) how the shape affects the wake parameters, such as $\alpha$,
$\lambda_{0}$ and $c$, and ultimately the $St-Re$ relationship.
Aside from its scientific interest, the inquiry is useful to a variety
of engineering problems where vortex shedding and wake formation play
an important role.

\begin{figure}
\begin{centering}
\includegraphics[width=8cm]{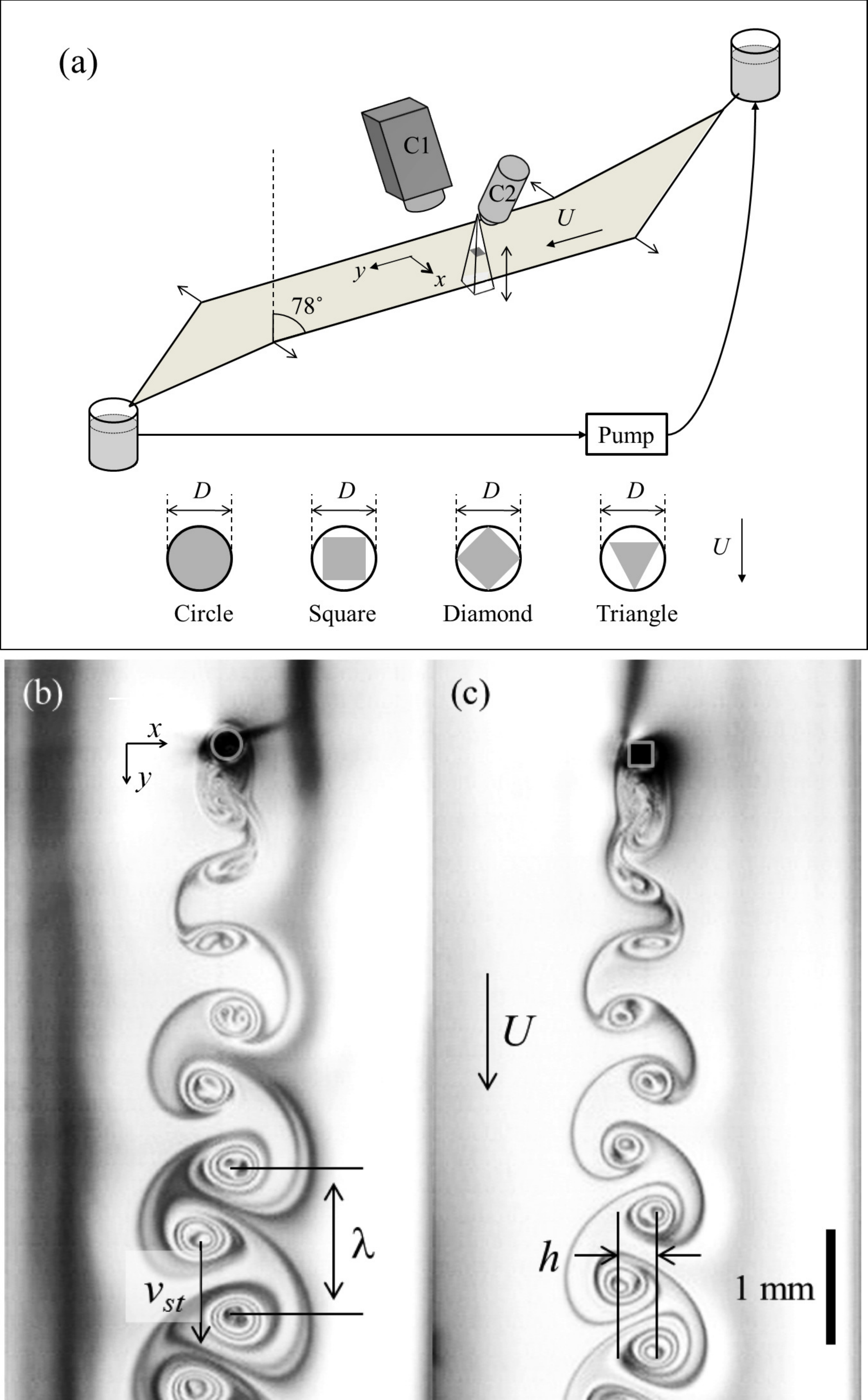}
\par\end{centering}

\protect\caption{(a) Experimental Setup. The soap film channel is inclined at $12\protect\textdegree$
from horizontal. A fast video camera C1 (Vision Research, Phantom
V5) and a microscope C2 (Wild M5A) equipped with a CMOS camera (DCM130,
Oplenic) are mounted directly above the film. Four different shaped
rods, circular (C), square (S), diamond (D), and triangular (T), are
used for measurements, and their characteristic size are defined as
$D$. Panels (b) and (c) display the vortex streets created by a circular
and a square rod, respectively. The Reynolds number, $Re\simeq170$,
is about the same for both cases. \label{fig:apparatus}}
\end{figure}

Our experiment was carried out in an inclined soap-film channel depicted
in Fig. \ref{fig:apparatus}(a) \cite{Rutgers_RevSciInstr_2001,Georgiev2002}.
The soap solution consisted of $2\%$ Dawn detergent, $5\%$ glycerol,
and water, giving a bulk kinematic viscosity $\nu\simeq0.013\,\text{cm}^{2}\text{/s}$.
The film was $2\,\text{m}$ long, $5\,\text{cm}$ wide, and flowed
continuously with a mean speed $U\simeq60\pm3\,\text{cm/s}$. At this
speed, the film has a thickness $\sim3\,\text{\ensuremath{\mu}m}$
as determined by a laser transmission method and is weakly compressible
with a Mach number $Ma\simeq0.12$ \cite{Kim2010}. Our flowing soap
film therefore served as a quasi-2D fluid medium with its slight compressibility
facilitating instantaneous flow visualization using a low-pressure
sodium lamp and a high-speed video camera. In this regard, the use
of a soap film is very attractive because the boundary layer separation
can be readily visualized without using dyes or other agents. The
physical basis of such flow visualization is that for a weakly compressibility
fluid, such as a soap film, the conserved quantity is $\omega/\rho_{2}$,
where $\omega$ is the vorticity and $\rho_{2}=\rho h$ is the 2D
density of the film. The variation of $\omega$ is thus accompanied
by a variation in $h$, since the water density $\rho$ is constant.
The technique is also very sensitive because the film thickness variation
$\Delta h$ is measured in terms of wavelength of the sodium lamp
(589 nm); a mere change of $\lambda/4$ in the thickness will cause
a change from constructive (bright) to destructive (dark) interference
in our video images. As Figs. \ref{fig:apparatus}(b,c) illustrate,
the technique allows direct measurements of flow structures, such
as the wavelength $\lambda$ and the street width $h$, and dynamic
parameters, such as $U$, $v_{st}$, and $f$. These measurements
were made without post processing of images. Here the shedding frequency
$f$ was determined by two methods: (i) directly counting the number
of vortices shed per second, and (ii) using the ratio $v_{st}/\lambda$
at a fixed downstream distance $y$. Both methods yields essentially
the same result as in Ref. \cite{Roushan2005a}. 

Vortex streets were created using tapered rods of different geometrical
cross sections, circle (C), square (S), diamond (D), and equilateral
triangle (T), as depicted in Fig. \ref{fig:alldata}(a-d). A circular
rod was made of glass using a glass puller. All other rods, including
another circular one, were made of titanium carefully machined to
have the tip size $\lesssim50\,\text{\ensuremath{\mu}m}$; the small
tips allow vortex street to be studied at small $Re$. Two circular
rods, made of glass and titanium, give identical results, suggesting
that the surface chemistry may not play a crucial role (also see Fig.
4.3(e) of Ref. \cite{Tuan2011}). To significantly reduce run-to-run
variations, we maintained a constant film thickness by keeping the
flow speed $U$ fixed. As delineated in Fig. \ref{fig:apparatus}(a),
$Re(\equiv UD/\nu)$ can be varied by changing the size $D$ of the
rod using a translation, and $D$ is measured by a CMOS camera mounted
on a long-working distance microscope.

\begin{figure}
\begin{centering}
\includegraphics[bb=20bp 10bp 732bp 572bp,clip,width=3.25in]{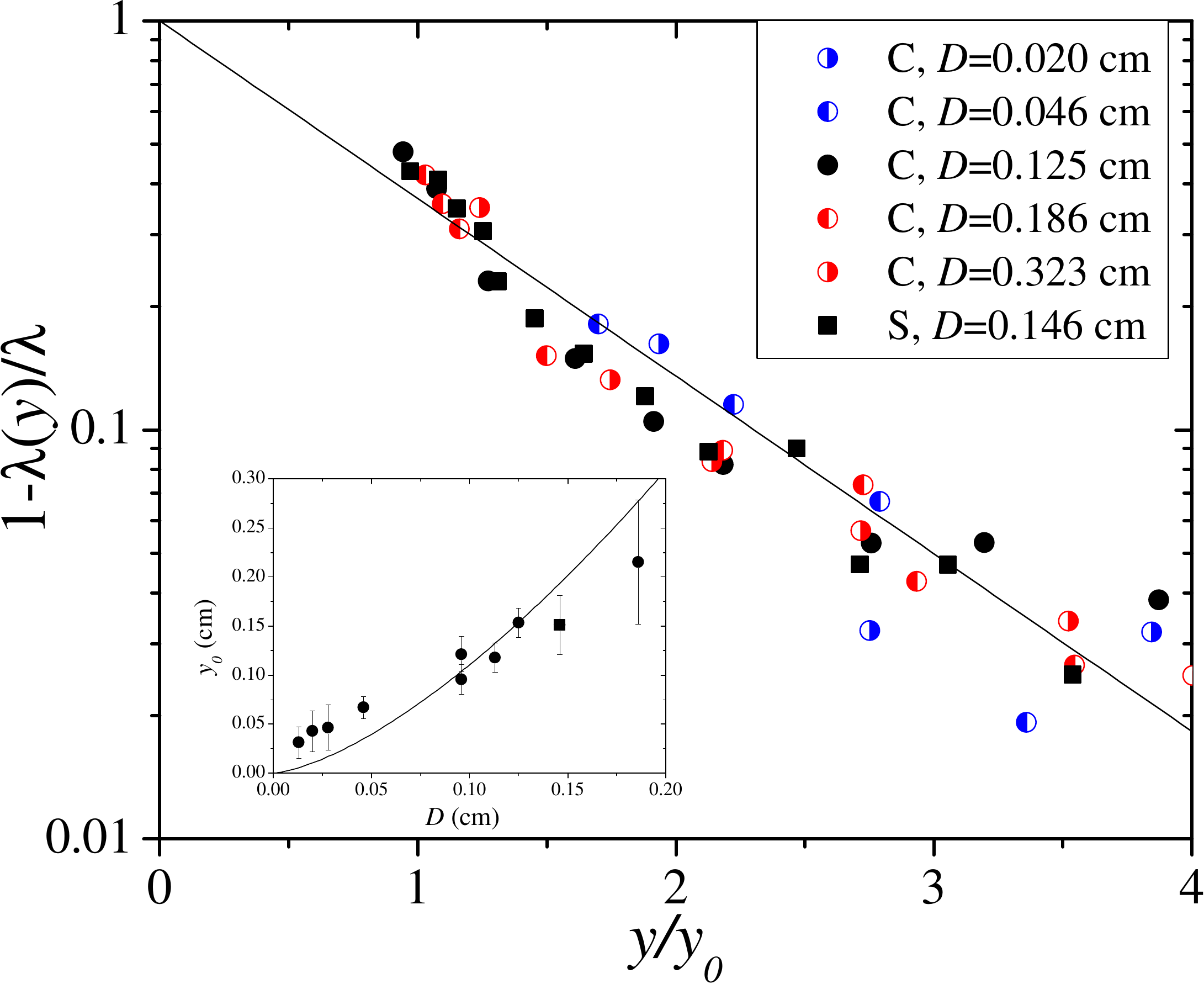}
\par\end{centering}

\protect\caption{(Color online) The downstream-distance dependent Wavelength $\lambda(y)$
vs. $y$. For all different rods, For circular (C) and square (S)
rods, the wavelength of their vortex streets depend on the downstream
distance $y$ and can be accurately described by $\lambda(y)=\lambda(1-e^{-y/y_{0}})$,
which is shown by the solid line. The inset shows that the decay length
$y_{0}$ depends on the size of the rods, $D$, and scales approximately
as $D^{3/2}$ as delineated by the solid line. This scaling behavior
is predicted by Eq. \eqref{eq:relaxation}. \label{fig:lambda_nondimensional} }
\end{figure}

As depicted in Figs. \ref{fig:apparatus} (b-c), directly beneath
the rods, vortices detach periodically from the rod, they then undergo
a transient relaxation over a downstream distance $y$, and finally
organize into a steady-state confirmation. The wavelength $\lambda(y)$
of a vortex streets therefore depends on $y$, and reaches a constant
value $\lambda$ for $y\gtrsim10D$. When $1-\lambda(y)/\lambda$
is plotted against $y,$ as depicted in Fig. \ref{fig:lambda_nondimensional}
for C and S rods, all the data for different $D$ follows a linear
behavior, suggesting an exponential dependence,

\begin{equation}
\lambda(y)=\lambda\left(1-e^{-y/y_{0}}\right),\label{eq:lambda(y)}
\end{equation}
where $y_{0}$ is the decay length. Systematic measurements, such
as this one, were carried out for the four rods, C, S, D, and T, and
their steady-state wavelengths $\lambda$ as a function of $D$ are
displayed in Figs. \ref{fig:alldata} (a-d). For all the cases we
found that $\lambda$, to a good degree, depends on $D$ linearly,
and the results of fitting using $\lambda=\lambda_{0}+\alpha D$ are
delineated by the red lines in the figures. Our experiment shows that
the intercepts $\lambda_{0}$ vary from rod to rod, but they are all
very small about a millimeter or so (see Table \ref{tab:fitting parameters}).
The slope $\alpha$ also depend on the shape of the rods with the
largest $\alpha=4.3\pm0.1$ for the C rod and the smallest $\alpha=3.1\pm0.1$
for the T rod.

\begin{figure*}
\begin{centering}
\includegraphics[bb=60bp 0bp 782bp 582bp,width=7in]{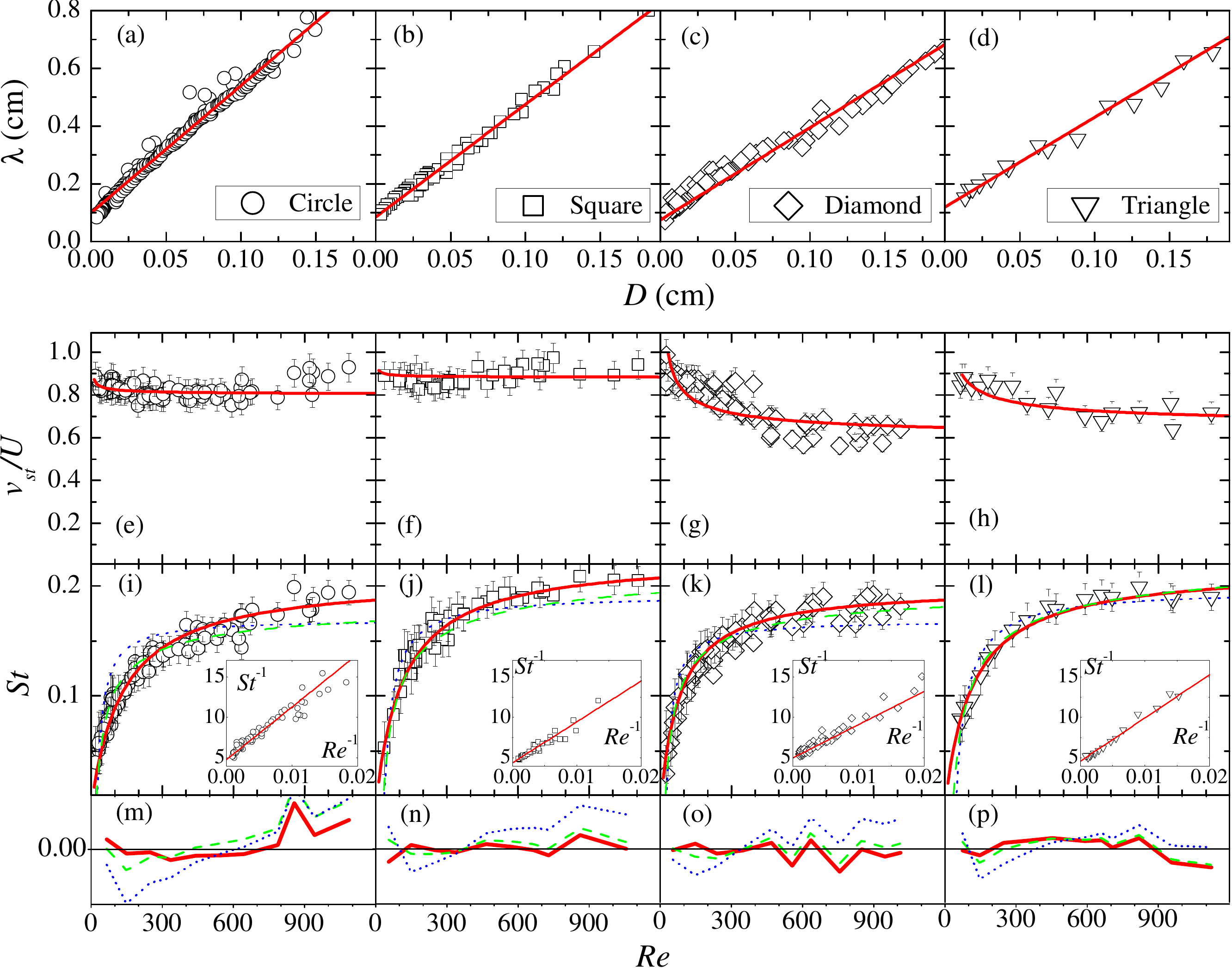}
\par\end{centering}

\protect\caption{(Color online) Experimental Results. The top panels (a-d) display
$\lambda$ vs. $D$ for circular (C), square (S), diamond (D), and
triangular (T) rods, respectively. In all the cases, $\lambda$ varies
linearly with $D$ but with a finite intercept $\lambda_{0}$ that
varies little among different rods. The middle panels (e-h) are $v_{st}/U$
vs. $Re$ for the same set of rods. For C and S rods, $v_{st}/U$
is approximately constant throughout the whole range of $Re$. However,
for D and T rods, $v_{st}/U$ decreases monotonically with $Re$ and
eventually reaches a plateau value defined as $c$. The bottom panels
(i-l) display $St$ vs. $Re$ for the four rods. The red solid line,
green dash line, and blue dotted lines in (i-l) are fitting curves
using respectively Eq. \eqref{eq:Roushan-Wu}, $St=a'-b'/\sqrt{Re}$
\cite{Fey1998,Williamson1998}, and the classical relation $St=a\left(1-b/Re\right)$.
The corresponding residuals of the fits are shown in (m-p). In the
insets of (i-l), the same graphs are replotted using $St^{-1}$ vs.
$Re^{-1}$. The linear relation suggests the validity of Eq. \eqref{eq:Roushan-Wu},
and moreover the intercepts, $St_{\infty}^{-1}\simeq4.6$, are about
the same for all the rods. \label{fig:alldata}}
\end{figure*}

In Figs. \ref{fig:alldata}(e-h), velocity of vortex streets relative
to the mean flow, $v_{st}/U$, is plotted as a function of $Re$.
Physically, the speed $v_{st}$ by which a vortex street travels in
the background flow $U$ depends on the vortex strength $\kappa$.
Since $\kappa$ is small near the onset, one expects $v_{st}/U\rightarrow1$,
i.e. vortices are passively convected by the mean flow. However, as
$Re$ increases and circulation in vortices becomes larger, one expects
$v_{st}/U$ to decrease. This qualitative behavior is indeed observed
for D and T, where $v_{st}/U$ decays monotonically as $Re$ increases
and levels off $v_{st}/U\rightarrow c$ for $Re>400$. Curiously,
this behavior is absent for C and S rods, where $v_{st}/U$ is nearly
constant for the whole range of $Re$; the quantity $v_{st}/U$ may
even increase slightly with $Re$, which results from mixing of vorticity
of opposite signs at very large $Re$. The plateau value $c$ is found
to depend on the shape of rods as detailed in Table with $c\simeq$0.81,
0.86, 0.63, and 0.70 for C, S, D, and T, respectively. The small $c$
values for D and T rods suggest that vorticity $\kappa$ is more efficiently
encapsulated into vortices by the rods with a trailing edge than rods
without it. Since in our experiment, the mean flow $U$ is fixed,
the above findings cannot be a result of air damping. The effect can
be understood, however, by the fact that a trailing edge reduces the
base suction pressure and keeps the two separated boundary layers
physically apart, reducing their mixing. This results in a wider wake
or a larger K\'{a}rm\'{a}n ratio, $K_{r}\equiv h/\lambda$, as will
be discussed later. 

It is useful at this point to compare our measurements with von K\'{a}rm\'{a}n's
point vortex model that makes predictions about the speed ratio of
the vortex street: $v_{st}/U=1-(\kappa/2U\lambda)\tanh(\pi h/\lambda)$
\cite{Birkhoff1953}. Assuming that K\'{a}rm\'{a}n's stability condition
\cite{Karman1911}, $\tanh(\frac{h\pi}{\lambda})=1/\sqrt{2}$ , holds
in the experiment and vorticity created in the boundary layer is $100\%$
encapsulated into the eyes of vortices, $\kappa=\lambda U$ \cite{Synge1927},
we found $v_{st}/U=1-\frac{1}{2\sqrt{2}}\simeq0.65$. This value is
remarkably close to the plateau value $c$ measured for D and T rods,
suggesting that these geometries permit nearly maximum preservation
of vorticity in the wake region. It also suggests that when $Re$
is not large or when objects do not have a trailing edge, such as
C or S rod, a noticeable amount of vorticity is annihilated before
a stable vortex street is formed. 

\begin{table}
\begin{centering}
\begin{tabular}{c||c|c|c|c}
\hline 
 & C & S & D & T\tabularnewline
\hline 
\hline 
$c$ & $0.81\pm0.04$ & $0.86\pm0.05$ & $0.63\pm0.05$ & $0.71\pm0.07$ \tabularnewline
\hline 
$\alpha$ & $4.3\pm0.1$ & $3.9\pm0.1$ & $3.2\pm0.1$ & $3.1\pm0.1$\tabularnewline
\hline 
$\lambda_{0}$ (mm) & $1.00\pm0.03$ & $0.85\pm0.06$ & $0.75\pm0.04$ & $1.2\pm0.1$\tabularnewline
\hline 
\hline 
$A$ & $5.1\pm0.2$ & $4.4\pm0.2$ & $5.0\pm0.2$ & $4.6\pm0.2$\tabularnewline
\hline 
$B$ & $580\pm32$ & $468\pm25$ & $456\pm29$ & $549\pm44$\tabularnewline
\hline 
\hline 
$\alpha/c$ & $5.3\pm0.3$ & $4.5\pm0.4$ & $5.1\pm0.5$ & $4.4\pm0.6$\tabularnewline
\hline 
$\lambda_{0}U/c\nu$ & $577\pm77$ & $456\pm81$ & $549\pm104$ & $791\pm200$\tabularnewline
\hline 
\end{tabular}
\par\end{centering}

\protect\caption{Measured Wake Parameters. For each shape (circle (C), square (S),
diamond (D), and triangle (T)) of the rod, parameters $c$, $\alpha$
and $\lambda_{0}$ are determined from Figs. \ref{fig:alldata}(a-h).
Coefficients $A$ and $B$ are determined using the plots in the insets
of Figs. \ref{fig:alldata}(i-l). For comparison, $\alpha/c$ and
$\lambda_{0}U/c\nu$ are also tabulated, and they provide an alternative
means to obtain $A$ and $B$ coefficients.\label{tab:fitting parameters}}
\end{table}

We now turn our attention to the $St$-$Re$ relation for different
rods. Here, the frequency $f$ was determined by counting the number
$N$ of vortices shed per second ($f=N/2$) and then non-dimensionalized
to obtain the Strouhal number, $St=fD/U$. We found that in all cases
laminar vortex streets persist over a wide range of $Re$, $10<Re<1200$,
which is in a sharp contrast with 3D measurements \cite{Williamson1988,Norberg1994,Zhang1994}.
As delineated in Figs. \ref{fig:alldata}(i-l), different rods exhibit
similar $St$-$Re$ dependence, i.e. $St$ increases rapidly for small
$Re$ and levels off for large $Re$. These behaviors can be accurately
captured by Eq. \eqref{eq:Roushan-Wu} as delineated by the solid
red lines in the figures. The appropriateness of Eq. \eqref{eq:Roushan-Wu}
is further checked by plotting $St^{-1}$ vs. $Re^{-1}$ as displayed
in the inset, where indeed good linear relationships are found. The
$A$ and $B$ coefficients extracted from these plots are listed in
Table \ref{tab:fitting parameters}, and overall they compare quite
well with those calculated based on structural measurements using
$A\equiv\alpha/c$ and $B\equiv\lambda_{0}U/c\nu$. The largest discrepancy
of $\sim30\%$ is for coefficient $B$ of D and T rods, and it is
due to approximating $v_{st}/U$ by a constant $c$, which according
to Figs. \ref{fig:alldata}(c) and (d) is valid only for large $Re$.
Surprisingly, even in these cases, parametrization of $St-Re$ relation
using only two parameters ($A$ and $B$) appears to be adequate based
on linearity of the data in the insets of Figs. (k-l). The most noteworthy
feature of these measurements is the fact that asymptotically ($Re\gg1$)
the Strouhal number $St_{\infty}\simeq0.21\pm0.02$ turns out to be
nearly the same for different rods. This suggests that $St_{\infty}$
(or $A$) is a property of the downstream wake rather than properties
of the obstacle that creates it. In Ref. \cite{Monkewitz1988}, it
has been proposed that street formation is a global instability of
the wake, and our measurement is consistent with this physical picture.
Our finding furthermore suggests that for blunt bodies, the mode selection
in the high $Re$ regime is independent of the body shape, indicating
that this mode may be universal.

\begin{figure}
\begin{centering}
\includegraphics{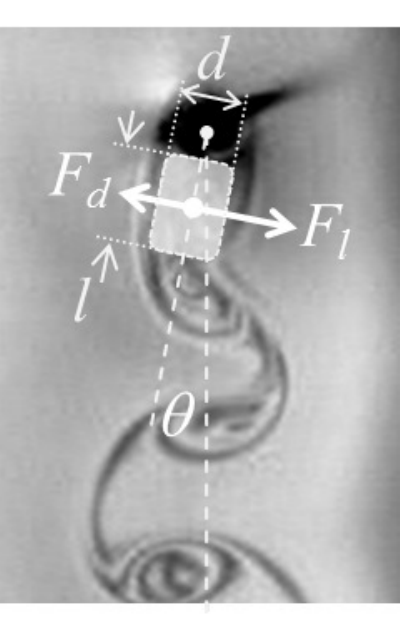}
\par\end{centering}

\protect\caption{The Free-Body Diagram Representing Birkhoff's Pendulum Model. Swing
of a fluid element in the near wake region is approximated by a physical
pendulum of length $\ell$ and width $d$ as indicated. The restoring
force is the lift indicated by $F_{l}$ and the damping force is indicated
by $F_{d}$.\label{fig:Free-body-diagram}}

\end{figure}

Inspection of a vortex street near a rod reveals a streak of fluid
that oscillates periodically in a fashion similar to a physical pendulum
(see Fig. \ref{fig:Free-body-diagram}). This observation was exploited
by Birkhoff to explain the experimentally observed $St$-$Re$ relationship
when $Re\gg1$ \cite{Birkhoff1953}. Below we generalize Birkhoff's
simple model to include viscous damping. As we shall see that with
such a modification, certain features of vortex streets observed in
our experiment can be described. For a lamina inclined at an angle
$\theta$ to the stream, it is well-known that the lift coefficient
is $C_{L}=2\pi\theta$ \cite{Landau}. This gives the cross-force
per unit length $F_{l}=\frac{1}{2}\rho U^{2}C_{L}=\pi\rho U^{2}\theta$.
For the fluid element of width $d$ and length $\ell$ behind the
rod, the inertia force per length is $\rho d(\ell\ddot{\theta)}$,
where $\rho$ is the 2D density. As for the damping term, the drag
force per unit length is $F_{d}=\frac{1}{2}\rho U^{2}C_{D}$, where
$C_{D}$ is the drag coefficient. For a cylinder, measurements showed
$C_{D}\propto1/\sqrt{Re}$ over a broad range of $Re$, $10\le Re\le10^{3}$
\cite{Schlichting}. It follows from a simple dimensional analysis
that the damping force per length is $\gamma_{0}\rho\nu\sqrt{Re'}\dot{\theta}$,
where $Re'=U\ell/\nu$ and $\gamma_{0}$ is a dimensionless constant
characterizing the overall magnitude of damping. Balancing these forces
yields,

\begin{equation}
\ddot{\theta}+2\tau_{0}^{-1}\dot{\theta}+\omega_{0}^{2}\theta=0.\label{eq:damped_oscillation}
\end{equation}
This equation describes the damped harmonic oscillations with a decay
time $\tau_{0}=2d\ell/\gamma_{0}\nu\sqrt{Re'}$ and a natural frequency
$\omega_{0}=U\sqrt{\pi/d\ell}$. If one assumes $\theta=\theta_{0}\exp\left(\Lambda t\right)$,
the characteristic value $\Lambda$ is given by, 
\begin{equation}
\Lambda=-\tau_{0}^{-1}\pm i\sqrt{\omega_{0}^{2}-\tau_{0}^{-2}}.\label{eq:dispersion_relation}
\end{equation}
Since the width of the wake is approximately the size $D$ of a rod,
we make an ansatz $d=D$ and $\ell=kD$, where $k$ is a phenomenological
parameter \cite{Birkhoff1953}. It follows from Eq. \eqref{eq:dispersion_relation}
that the oscillation frequency $f$ of the wake is given by, 

\begin{eqnarray}
f=\frac{{\rm Im}(\Lambda)}{2\pi} & = & \frac{U}{2\sqrt{\pi d\ell}}\left(1-\frac{\gamma_{0}^{2}\nu}{4\pi Ud}\right)^{1/2}\nonumber \\
 & \simeq & \frac{U}{2D\sqrt{k\pi}}\left(1-\frac{\gamma_{0}^{2}}{4\pi Re}\right)^{1/2},\label{eq:shedding_frequncy}
\end{eqnarray}
where $Re\equiv UD/\nu$. In the small-damping limit ($\gamma_{0}^{2}/(4\pi Re)\ll1$),
Eq. \eqref{eq:shedding_frequncy} gives $St^{-1}(\equiv U/fD)$= $St_{\infty}^{-1}(1+\gamma_{0}^{2}/8\pi Re)$,
which has the same mathematical form as the phenomenologically derived
$St$-$Re$ relation, Eq. \eqref{eq:Roushan-Wu}. Here $St_{\infty}\equiv1/2\sqrt{k\pi}$
is the asymptotic Strouhal number and is identical to Birkhoff's result
\cite{Birkhoff1953}. Since $St_{\infty}\simeq0.2$ is nearly a constant
for different rods (see Fig. \ref{fig:alldata}(i-l)), it may be concluded
that $k=2$ and $St_{\infty}=1/2\sqrt{2\pi}$ is universal for a laminar
vortex wake. We also notice that in the same small-damping limit,
Eq. \eqref{eq:shedding_frequncy} yields the experimentally observed
linear $D$ dependence for $\lambda$, $\lambda=\lambda_{0}+\alpha D$,
where $\alpha=2\sqrt{2\pi}c$ and $\lambda_{0}=c\gamma_{0}^{2}\nu/2\sqrt{2\pi}U$.
For a circular rod and using $c=0.81$ in Table \ref{tab:fitting parameters},
we found $\alpha\simeq4.1$, which is in reasonable agreement with
the slope $\alpha\simeq4.3$ seen in Fig. \ref{fig:alldata}(e). 

Finally, the real part of Eq. \eqref{eq:dispersion_relation} gives
the characteristic relaxation time $\tau_{0}$ of the oscillation,
\begin{equation}
\tau_{0}=\frac{1}{{\rm Re}(\Lambda)}=\frac{2\sqrt{kRe}}{\gamma_{0}}\frac{D}{U}.\label{eq:relaxation}
\end{equation}
If one associates $\tau_{0}$ with the decay length $y_{0}$ of the
wake defined in Eq. \eqref{eq:lambda(y)}), i.e. $y_{0}\simeq v_{st}\tau_{0}$,
Eq. \eqref{eq:relaxation} suggests the scaling $y_{0}\propto D^{3/2}$.
In the inset of Fig. \ref{fig:lambda_nondimensional}, this predicted
relationship (solid line) is compared with the measured decay length
$y_{0}$ vs. $D$ for C rod. The agreement is fair considering that
uncertainties in the measurement is quite large. 

In summary, we showed that the phenomenologically derived $St$-$Re$
relation, Eq. \eqref{eq:Roushan-Wu}, is applicable to vortex shedding
behind blunt bodies other than circular ones. Specifically, the $A$
and $B$ coefficients in the equation are determined respectively
by two characteristic length scales $D$ and $\lambda_{0}$ in the
flow. A significant finding of this work is that in the high $Re$
regime, the wake oscillation frequency $f$ is uniquely determined
by the largest length scale $D$ in the problem, resulting in $St_{\infty}\rightarrow1/A\simeq0.21$
(or $f\simeq0.21U/D$) for all different rods. On the other hand,
in the low and intermediate $Re$ regimes, where the fluid viscosity
cannot be neglected, $\lambda_{0}$ also contributes to vortex shedding,
and $St$ becomes shape dependent. There exists a strong correlation
between the street velocity $v_{st}$, characterized by $c=v_{st}/U$
at large $Re$, and the shape of a body, e.g., for bodies with a trailing
edge, such as D and T rods, $c$ are significantly lower than that
of C and S rods. Since $c$ is a measure of the vortex strength $\kappa$,
it can be concluded that the trailing edge allows more powerful vortices
to be shed and better preserved downstream. We noticed moreover that
when $c$ is small, the wake parameter $\alpha$ is also small. This
gives rise to interesting properties of a wake, such as $St_{\infty}=1/A\simeq c/\alpha$
being weakly shape dependent but the K\'{a}rm\'{a}n ratio $K_{r}=h/\lambda$
strongly shape dependent. The latter can be seen by noticing that
since $h\simeq D$ and $\lambda=\lambda_{0}+\alpha D\simeq\alpha D$,
the K\'{a}rm\'{a}n ratio is given by $K_{r}\simeq\alpha^{-1}$.
For C, S, D, T rods, we obtained $K_{r}\simeq0.23,\,0.25,\,0.31,$
and $0.32$, respectively. In K\'{a}rm\'{a}n's classical calculation,
it was found that point vortex street is stable when $K_{r}=\frac{1}{\pi}\cosh^{-1}\sqrt{2}\simeq0.28$.
Our experiment shows that vortex streets generated with different
shaped rods have $K_{r}$ not exactly as K\'{a}rm\'{a}n had predicted,
but interestingly they all appear to be stable.

This work is supported by the NSF under the grant no. DMR-0242284.

\bibliographystyle{apsrev4-1}
\bibliography{vortex2012}

\end{document}